\documentclass[12pt]{article}

\newcommand{\blind}{0}

\usepackage[authoryear, round, longnamesfirst]{natbib}
\usepackage[english]{babel}
\usepackage[T1]{fontenc}

\usepackage{amsmath}
\usepackage{amsthm}
\usepackage{amssymb}

\usepackage[dvipsnames]{xcolor}
\usepackage[colorlinks=true, linkcolor=MidnightBlue, citecolor=RoyalBlue, urlcolor=darkgray]{hyperref}
\usepackage{graphicx}
\usepackage{adjustbox} 
\usepackage{float}     
\usepackage[section]{placeins}
\usepackage{subfig}
\usepackage{listings}
\usepackage{multirow}	
\usepackage{titling}

\setkeys{Gin}{width=0.8\textwidth}

\usepackage{orcidlink}

\def\spacingset#1{\renewcommand{\baselinestretch}%
{#1}\small\normalsize} \spacingset{1}

\begin{document}

\title{\bf Variable Importance in Generalized Linear Models -- A Unifying View Using Shapley Values
}

\medskip

\if0\blind
{
\author{Sinan Acemoglu\thanks{Universit\"at Basel, Basel, Switzerland. Email: sinan.acemoglu@unibas.ch}
\and
Christian Kleiber\orcidlink{0000-0002-6781-4733}\thanks{Universit\"at Basel, Basel, Switzerland. Email: christian.kleiber@unibas.ch}
\and
J\"{o}rg Urban\thanks{Universit\"at Basel, Basel, Switzerland. Email: joerg.urban@unibas.ch}}
} \fi

\if1\blind
{
\author{}
} \fi

\date{\today}

\maketitle

\thispagestyle{empty}

\begin{abstract}
Variable importance in regression analyses is of considerable interest in a variety of fields. There is no unique method for assessing variable importance. However, a substantial share of the available literature employs Shapley values, either explicitly or implicitly, to decompose a suitable goodness-of-fit measure, in the linear regression model typically the classical $R^2$. Beyond linear regression, there is no generally accepted goodness-of-fit measure, only a variety of pseudo-$R^2$s. We formulate and discuss the desirable properties of goodness-of-fit measures that enable Shapley values to be interpreted in terms of relative, and even absolute, importance. We suggest to use a pseudo-$R^2$ based on the Kullback-Leibler divergence, the Kullback-Leibler $R^2$, which has a convenient form for generalized linear models and permits to unify and extend previous work on variable importance for linear and nonlinear models. Several examples are presented, using data from public health and insurance.

\medskip


\end{abstract}

\newpage

\spacingset{1.45} 

\section{Introduction}
\label{Intro}

Assessing the importance of explanatory variables in regression analyses is of considerable interest in a number of areas. There is a large literature spanning several decades, highly fragmented and characterized by parallel developments and rediscoveries in numerous fields, among them various behavioral and social sciences, the medical sciences, ecology, and business administration. Excellent surveys are available from \citet{Azen+Budescu:2003}, \citet{Johnson+Lebreton:2004}, and \citet{Groemping:2007,Groemping:2015}.

For the linear regression model, a prominent method, apparently originating from \citet{Lindeman+Meranda+Gold:1980}, determines variable importance via a representation of the classical $R^2$ as a weighted average involving all possible subsets of regressors in the model. At the time it did not attract much attention, presumably because of the substantial computational burden. (Note that a model with 10 regressors -- a model of rather modest size by current standards -- already involves $2^{10}=1024$ regressions.) The beauty of the proposal in \citet{Lindeman+Meranda+Gold:1980} is that it is easy to interpret and that it produces a `fair' decomposition of the value of an overall goodness-of-fit measure, $R^{2}$, into the contributions of the individual regressors. The method has been rediscovered and extended several times using different terminologies, among them {\it{averaging over orderings}} \citep[a term that we will use below, see][]{Kruskal:1984, Kruskal:1987}, {\it{hierarchical partitioning}} \citep{Chevan+Sutherland:1991} and {\it{dominance analysis}} \citep{Budescu:1993, Azen+Budescu:2006}. More specifically, hierarchical partitioning and dominance analysis extend the idea of \citet{Lindeman+Meranda+Gold:1980} by permitting fairly arbitrary goodness-of-fit measures for evaluating the contributions of predictors. They also go beyond linear regression. 

It was pointed out by \citet{Stufken:1992} that hierarchical partitioning, and hence averaging over orderings, amounts to using the Shapley value, a concept from cooperative game theory, to decompose a measure of fit of the regression. An early paper explicitly using the Shapley value framework is \citet{Lipovetsky+Conklin:2001}, still confined to linear regression. 

Advances in statistical computing and increasing computational power have made Shapley value computations more feasible in recent years, and the availability of flexible software for nonlinear models has motivated the search for measures of variable importance beyond linear regression. Indeed, \citet{Chevan+Sutherland:1991} already suggested to use hierarchical partitioning to measure the contributions of predictors in classical nonlinear models via a decomposition of $\chi^{2}$ statistics. Later, \citet{Azen+Traxel:2009} applied dominance analysis in logistic regression, but now with certain pseudo-$R^{2}$ measures. To the best of our knowledge, empirical examples for nonlinear models are still largely confined to binary response models, although some software implementations offer functionality beyond this setting, notably Poisson regression. However, there are problems with certain proposals and corresponding implementations, as we shall discuss below.

In this paper, we pursue three goals: (i)~the formulation of desirable properties of goodness-of-fit measures that allow an interpretation of Shapley values in terms of relative and even absolute importance, (ii)~the systematic use of Shapley values for a reasonably large class of models, here GLMs and certain extensions thereof, and (iii)~the consistent use of Shapley values with goodness-of-fit measures that generalize classical variants while at the same time opening the methodology to further settings. 

We show that the goodness-of-fit measures that are used to assess the contributions of predictors need to be chosen carefully, as their properties are critical to the interpretation of the resulting Shapley values. We therefore formulate desirable properties of goodness-of-fit measures that help to interpret the resulting Shapley values as measures of relative and even of absolute importance. 
Currently, there does not appear to be a generally accepted procedure for assessing variable importance in generalized linear models (GLMs). Therefore, we propose to use Shapley values along with a fit measure that meets certain requirements and exists for all GLMs, namely the Kullback-Leibler $R^{2}$, hereafter denoted as $R^{2}_{\text{KL}}$. It was introduced by \citet{Cameron+Windmeijer:1997} and is closely related to the deviance and the classical likelihood ratio statistic. 
Specifically, for the linear regression model $R^{2}_{\text{KL}}$ reduces to the classical $R^2$, while for binary response models it reduces to McFadden's likelihood ratio index, perhaps the most prominent of all pseudo-$R^2$ measures in this setting. Thus, our proposal extends Shapley value decompositions from linear regression models to the much larger class of GLMs (and even some related models), thereby providing a unified approach to variable importance for a range of nonlinear regression models. We present several empirical examples to illustrate the methodology, using data from public health and insurance. 

The remainder of this paper is organized as follows: In Section~\ref{SV} we summarize the necessary background on the Shapley value. In Section~\ref{ShvIntOv} we discuss desirable properties of a goodness-of-fit measure for which Shapley values are computed. 
We establish that $R^{2}_{\text{KL}}$ possesses these properties in Section~\ref{R2KL}. Section~\ref{ShvExample} provides examples involving Poisson and geometric regression as well as the Poisson hurdle model, a model with two linear predictors but with GLM building blocks. These examples also illustrate the tradeoffs in the choice of the fit measure. Section~\ref{conclusion} concludes.

\section{Shapley values in regression}
\label{SV}
The Shapley value is a solution concept from cooperative game theory, introduced by \citet{Shapley:1953}. A convenient reference for the game-theoretical terminology and background is \citet{Ferguson:2020}. A cooperative game $(P,v)$ in coalitional form is described by a finite set of players, $P = \{1,\dots,p\}$, and a characteristic function $v: 2^P \to \mathbb{R}$ that assigns a real number $v(S)$ to each element $S$ of the power set $2^P$. $S \subseteq P$ is called a coalition, $P$ the grand coalition, and $v(S)$ can be interpreted as the payoff that coalition $S$ can secure when its members act as a unit. In cooperative game theory, a standard condition for the characteristic function $v$ is $v(\varnothing) = 0$; i.e., the empty set or coalition $\varnothing$ secures a payoff of zero. We may refer to this as `zero-normalization'.


%

The Shapley value $\varphi_i(P,v)$ for player $i \in P$ and a given characteristic function $v$ is 
\begin{equation}
\label{shapleyvalue}
\varphi_i(P,v) 
~=~ \sum_{S \subseteq P \setminus \{i\}}^{} 
\underbrace{ \frac{|S|! ~ (p-|S|-1)!}{p!}}_{\rm{weight} }
\underbrace{ \left( v(S \cup \{i\}) - v(S) \right) }_\text{\parbox{4cm}{\centering marginal contribution of\\[-4pt] player $i$ to coalition $S$}},
\end{equation}
where $|S|$ represents the cardinality of the set $S$. Hence $\varphi_i(P,v)$ is the average marginal contribution of player~$i$; the average is formed over all subsets $S$ of $P$ that do not contain player~$i$. 
Following \citet{Kruskal:1987}, one may call this approach \emph{averaging over orderings}.

Among the various properties of the Shapley value, the \textit{efficiency property} \citep{Shapley:1953}
\begin{equation}\label{shv_eff}
\sum_{i\in P} \varphi_i(P,v) ~=~ v(P)
\end{equation}
is of particular importance for this paper. It states that the value of the characteristic function evaluated at the entire set of players, the grand coalition $P$, is identical to the sum of all Shapley values. 

\subsection*{Shapley values in linear regression}

As noted above, Shapley values emerged implicitly in the variable importance literature via the principle of averaging over orderings. In a regression context, the set of `players' is a set  $P=\{1,2,\dots,p\}$ of regressors that is used to predict a particular outcome. It leads to $2^p$ different models defined by the different subsets of regressors that can be formed. The role of the characteristic function $v$ is played by a suitable goodness-of-fit measure. In the linear regression model typically the classical $R^2$ is used, which satisfies $v(\varnothing) = 0$. 
In view of the efficiency property (\ref{shv_eff}), each Shapley value $\varphi_i(P,v)$ can be interpreted as the contribution of regressor $i$ to the model's overall $R^2$, hence in this sense it measures the regressor's relative importance. 

\citet[][Sec.~4.7]{Lindeman+Meranda+Gold:1980} suggest a measure of importance based on semi-partial correlations $r_{(i\cdot S)}^{2}$. These semi-partial correlations $r_{(i\cdot S)}^{2}$ measure the correlation between the response $y$ and the regressor $i$, with the correlations of the other predictors in $S \subseteq P$ partialed out. The measure can be expressed as a weighted average over the increments in $R^2$ resulting from the inclusion of predictor~$i$, specifically, with $v = R^2$ 

\begin{eqnarray}
\label{shv_semipart}
\varphi_i(P, R^{2}) 
&=& \sum_{S \subseteq P \setminus \{i\}}^{}  \frac{|S|! ~ (p-|S|-1)!}{p!} \, r_{(i\cdot S)}^2 \nonumber \\ 
&=& \sum_{S \subseteq P \setminus \{i\}}^{}  \frac{|S|! ~ (p-|S|-1)!}{p!} \, \left( R^2(S \cup \{i\}) - R^2(S) \right).
\end{eqnarray}
Here $R^2(S)$ and $R^2(S \cup \{i\})$ correspond to the $R^{2}$s of the models whose sets of predictors are $S$ and $S \cup \{i\}$, respectively. Thus \citet{Lindeman+Meranda+Gold:1980} have implicitly proposed the use of Shapley values in regression.

Applications to linear regressions abound; \citet{Groemping:2007,Groemping:2015} provides many references. 

\subsection*{Shapley values beyond linear regression}

\citet{Chevan+Sutherland:1991} already suggested to use Shapley values beyond the linear regression model, explicitly mentioning logistic, probit, and log-linear regression. Within the framework of hierarchical partitioning, \citet{Walsh+Papas+Crowther:2004} calculated Shapley values for the logit model using the log-likelihood as the characteristic function. Within the framework of dominance analysis, \citet{Azen+Traxel:2009} provide a further application to binary response models. To overcome rescaling issues inherent in the use of the likelihood, they suggest the use of certain pseudo-$R^{2}$ measures to calculate Shapley values. Among these pseudo-$R^2$ measures, they express a slight preference for the likelihood ratio index of \citet{McFadden:1973}, hereafter denoted as $R^{2}_{\text{McF}}$. More recent applications include \citet{Yu+Zhou+Suh:2015} and \citet{Lee+Dahinten:2021}. 
\citet{Nandintsetseg+Shinoda+Du:2018} apply dominance analysis to Poisson regression using $R^{2}_{\text{McF}}$, and \citet{Tetteh+Ekem-Ferguson+Quarshie:2021} also evaluate variable importance for the Poisson model through Shapley values.

Applications of dominance analysis even beyond GLMs exist, see \citet{Shou+Smithson:2015} for an example using beta regression. Noting that pseudo-$R^{2}$ measures such as McFadden's likelihood ratio index $R^{2}_{\text{McF}}$, originally designed for binary data, may not be appropriate in their setting (which involves a continuous distribution), they use characteristic functions such as the BIC and the likelihood ratio test statistic. Furthermore, an application involving two-part models, also known as hurdle models, is sketched in \citet{Lima+Ferreira+Leal:2021}. Their two-part model consists of two GLM building blocks, a logit model and a gamma regression model. We also provide an example of a two-part model in Section \ref{HPM} below, where we consider the widely used Poisson hurdle model. 

\citet{Shou+Smithson:2015} emphasize that the choice of the goodness-of-fit measure is important in nonlinear settings. Indeed, in Section \ref{ShvIntOv} we show that a careful choice of the fit measure is essential for the validity of the efficiency property~(\ref{shv_eff}), and, consequently, for the interpretation of the resulting Shapley values and the unification of the associated methodology. 

We observe that there are numerous potential choices for the characteristic function $v$. For example, the machine learning (ML) literature has recently shown a trend towards `explainable' or `interpretable' ML, in which Shapley values are employed to decompose predictions or prediction errors. Influential papers in this area include \citet{Strumbelj+Kononenko:2010} and \citet{Lundberg+Lee:2017}, which contain references to earlier work in computer science and related fields. In a predictive setting where $v$ represents a conditional expectation, neither $v(\varnothing) = 0$ nor monotonicity with respect to the addition of further predictors is generally satisfied. Therefore, the resulting Shapley-based decomposition does not represent a decomposition of a fit measure. In contrast, in line with earlier developments in the statistical literature, we focus on decomposing a goodness-of-fit measure. This requires a characteristic function with certain monotonicity properties. More on this in Section~\ref{ShvIntOv}. 

\subsection*{Software}

Dominance analysis and hierarchical partitioning have been implemented in several \textsf{R} packages, permitting to decompose quantities such as the log-likelihood or the pseudo-$R^{2}$ measures used by \citet{Azen+Traxel:2009}. The hierarchical partitioning procedure is available from the \textbf{hier.part} package \citep{Mac-Nally+Walsh:2004} and dominance analysis from the package \textbf{dominanceanalysis} \citep{Navarrete+Soares:2020}. For linear models, the package \textbf{relaimpo} \citep{Groemping:2006} also provides methods that are not derived from the principle of averaging over orderings.

\section{The role of the goodness-of-fit measure}
\label{ShvIntOv}
In applications, it is desirable that measures of variable importance are easy to interpret. 
In this section, we show that under certain conditions this goal can be achieved when using Shapley values. We next discuss desirable properties of the goodness-of-fit measure (the characteristic function in the original game-theoretical context).  

\subsection{Monotonicity}
\label{monotonicity}
From a regression point of view, it is natural to require a characteristic function that is weakly increasing when a new predictor is added, i.e., $v(S \cup \{i\}) \geq v(S)$ for any $S \subseteq P$ and $i \in P \setminus S$. This is a meaningful condition because we expect a goodness-of-fit measure to improve, or at least to remain unchanged, when a new explanatory variable is added. In view of (\ref{shapleyvalue}), the resulting Shapley values are nonnegative under this condition. 

 
\subsection{Lower bound}
\label{LowerBound}
\citet{Shapley:1953} requires a characteristic function to satisfy $v(\varnothing) = 0$ from Section \ref{SV}. In a regression context, this condition means that the fit of the model not using any regressors (beyond a constant term) is zero. 
The condition $v(\varnothing) = 0$ along with monotonicity also results in $v(S) \geq 0$ for all $S \subseteq P$. We call this the lower bound condition.


In our context, the set of Shapley values $\varphi_{i}(P, v)$ represents a decomposition of the fit measure evaluated at the full set of regressors, $\sum_{i \in P} \varphi_{i}(P, v) = v(P)$. Hence the (normalized) Shapley values can be interpreted as shares relative to the fitted model, $FM$; i.e., we can define an importance measure for variable $i$ by setting
\begin{align}
\label{ShvDecomp_imp}
impFM_i 
~:=~ \dfrac{\varphi_{i}(P, v)}{\sum_{j \in P} \varphi_{j}(P, v)} 
~=~ \dfrac{\varphi_{i}(P, v)}{v(P)}, 
\qquad {\rm with} \; \sum_{i=1}^p impFM_{i} ~=~ 1. 
\end{align}
For later use, we briefly explore the implications for interpretability when the zero-normalization condition for the characteristic function is not satisfied. Specifically, consider a `pseudo-characteristic function' $v^*$ with $v^*(\varnothing)\neq 0$ and denote the resulting `pseudo-Shapley values' by $\varphi^{*}_i$. Next, define $v$ by $v(\cdot) = v^*(\cdot) - v^*(\varnothing)$, which represents a zero-normalized characteristic function. Starting from equation (\ref{shapleyvalue}), we see that, by construction,
\begin{equation}
\label{pseudoShapley}
\varphi^{*}_i(P,v^*) ~=~ \varphi_i(P,v),
\end{equation}
because the building blocks of Shapley values are differences of $v^*$ for `pseudo-Shapley values' and differences of $v$ for Shapley values. Therefore, we can also use the `pseudo-Shapley values' to establish a ranking of the predictors. Specifically, if $\varphi^{*}_{A} > \varphi^{*}_{B}$, then predictor $A$ is more important than predictor $B$ within the fitted model. 

Furthermore, using equation (\ref{pseudoShapley}) and the efficiency property, we have  
\begin{equation}
\label{ShvFin1}
\sum_{i} \varphi^{*}_i(P,v^*) ~=~ \sum_{i} \varphi_i(P,v) ~=~ v(P) ~=~ v^*(P) -v^*(\varnothing) ~\neq~ v^*(P).
\end{equation}
Thus, a violation of the zero-normalization condition for the characteristic function has the implication that the implied pseudo-Shapley values do not correspond to the predictors' contributions to the overall $v^{*}(P)$ of the fitted model because the efficiency property does not hold for $\varphi^{*}_i(P,v^*)$. Indeed, the resulting pseudo-Shapley values sum up to $v(P)$ and should therefore be interpreted with respect to $v(P)$ and not $v^*(P)$.  
Furthermore, the quantities $impFM_i$ based on pseudo-Shapley values do not correspond to the relative contribution of the overall explanatory power $v^{*}(P)$ of the fitted model; i.e., they do not add up to 1. This is a consequence of the shift term $v^{*}(\varnothing)$ in equation (\ref{ShvFin1}). 
It follows that for interpretational purposes pseudo-Shapley values are of limited usefulness and should only be used for ordinal comparisons of predictors. We will return to this issue in Section \ref{ShvExample}, where we will illustrate problems arising from the use of a prominent example of a pseudo-characteristic function, the log-likelihood function~$\ell$. The log-likelihood $\ell$ generally does not satisfy the zero-normalization condition, so it will lead to pseudo-Shapley values instead of genuine Shapley values, which raises various interpretational issues.

\subsection{Upper bound}
\label{UpperBound}
Recall that the empty set corresponds to a model with no regressors and thus describes the `worst' model. Analogously, the `best' model could be defined in a data driven manner, where each observation is given its own regression coefficient. This is called the \emph{saturated model} in a GLM setting \citep[e.g.][p.~274]{Dunn+Smyth:2018} and corresponds to the model for which the likelihood is maximized for a given set of data and a given type of model \citep[][p.~33]{McCullagh+Nelder:1989}. Therefore, it represents a further suitable reference point in our context. 

Just as in the case of the `worst benchmark model' discussed above, problems of interpretation can also arise when a goodness-of-fit measure $v$ is used that cannot be interpreted relative to some `best benchmark model'. More formally, suppose $v: S \rightarrow [0, b]$, $b\in\mathbb{R}_+${\footnote{Due to the monotonicity and the zero-normalization condition, the range of values of the goodness-of-fit measure $v$ is a subset of $\mathbb{R}_+$.} and $S \subseteq P'$. Here $P'$ represents the saturated model. Then, $v(\varnothing) = 0$ and $v(P') = b$ represent the evaluations of the `worst' and `best' models, respectively. 

In the original \citet{Shapley:1953} setting, there is no upper bound on the range of values of the characteristic function. However, in a regression context the overall fit $v(P)$ is difficult to interpret in the absence of an upper bound. This is mainly due to the fact that the value $b$ of the goodness-of-fit measure of the `best' baseline model may be unknown or unavailable (e.g., in the case of the log-likelihood function $\ell$). Here, we can still use  $\varphi_i(P,v)$ to assess relative importance similar to Subsections \ref{monotonicity} and \ref{LowerBound}. 
However, the implications for interpretability remain, because a seemingly large Shapley value $\varphi_i(P,v)$ may indicate great importance while the overall fit of the model may be poor. The poor fit would not be recognized unless $b$ is known. In other words, a large Shapley value could still correspond to a predictor of limited relevance.

This problem can be overcome if the fit of the `best benchmark model' is known. Then, the Shapley value can be interpreted as the importance relative to the best model, $BM$,  
\begin{align}
\label{ShvDecomp_impBest}
impBM_i ~:=~ \dfrac{\varphi_{i}(P, v)}{b}. 
\end{align}
If we additionally have a characteristic function for which the finite upper bound is equal to unity, i.e.,  $b=1$, the Shapley values can now even be interpreted as absolute importance measures relative to the best model achievable. Furthermore, a comparison is now possible for all models that are estimated using the same model class and data. To avoid the problems discussed above, we therefore suggest using a goodness-of-fit measure with an upper bound of one. 

\subsection{Structural interpretability of the fit measure}
\label{InterpretGLM}
So far, we have discussed conditions that ensure a straightforward interpretation of Shapley values as importance measures, leading to the ability to rank regressors and to assess whether their contributions are large relative to the fitted model or to some `best' model. 

However, in addition to the points made in previous subsections, the interpretation of the resulting Shapley values depends highly on the structure of the fit measure itself. For example, the classical $R^{2}$ in the linear regression model corresponds to the fraction of the overall variance that is explained by the model. Thus, when using the classical $R^{2}$ in a linear model, the Shapley value for a predictor can be interpreted as the share of the variance that is explained by this predictor \citep{Lindeman+Meranda+Gold:1980}. In contrast, using a likelihood-based quantity as the goodness-of-fit measure, for example, does not usually result in a variance decomposition beyond the linear regression model. 

Thus, although the conditions from Subsections \ref{monotonicity} -- \ref{UpperBound} already permit interpretation of Shapley values at a certain level, a suitable choice of the goodness-of-fit measure can lead to additional insights. Therefore, we suggest using a goodness-of-fit measure that has a meaningful structural interpretation for a range of regression models, such as GLMs. A suitable candidate is the Kullback-Leibler $R^2$, whose properties are summarized in Section~\ref{R2KL}.

\subsection{Desirable properties of the fit measure}
\label{Req}
The previous subsections have provided insights into the importance of the choice of the goodness-of-fit measure as the characteristic function, its structure, interpretability and the implications thereof. In view of the problems  mentioned above, the following properties are desirable:
\begin{itemize}
\item[(i)] Monotonicity: $v$ is (weakly) non-decreasing when a new predictor is added,
\item[(ii)] Lower bound: $v(\cdot) \geq 0$,
\item[(iii)] Upper bound: $v({\text{saturated model}}) = 1$,
\item[(iv)] Structural interpretability of $v$ for GLMs.
\end{itemize}
The monotonicity condition (i) ensures that all resulting Shapley values are non-negative and are comparable with each other through relative orderings. In addition, condition (ii) ensures that a Shapley value can be interpreted as contribution to the overall fit $v(P)$ and thus as the importance of the  variable \emph{relative to the fitted model}. Condition (iii) extends this interpretation even to importance  \emph{relative to the best achievable model}, i.e., to absolute importance. Table \ref{tab:cond_versus_properties} summarizes these interrelations.
\begin{table}[H]
\begin{center}
\caption{Connections among the properties of the fit measure and the properties of the corresponding Shapley and pseudo-Shapley  values} 
\begin{adjustbox}{width=\textwidth}
\begin{tabular}{rcccc}
 \hline 
 \hline 
  & \multicolumn{2}{c}{pseudo-Shapley values as}     & Shapley values as            & Shapley values as relative and\\
  & \multicolumn{2}{c}{relative variable importance} & \hphantom{XX} relative variable importance\hphantom{XX} & absolute variable importance \\
\hline
  & nonnegative     & relative & importance relative & importance relative \\
Condition: & contribution & ordering & to fitted model $impFM$      & to `best' model $impBM$ \\
\hline
 (i) Monotonicity  & $\times$ & $\times$ & $\times$ & $\times$\\
 (ii) Lower bound  &          &          & $\times$ & $\times$ \\ 
 (iii) Upper bound &          &          &          & $\times$
\end{tabular}
\end{adjustbox}
\label{tab:cond_versus_properties}
\end{center}
\end{table}
Therefore, if conditions (i)--(iii) are satisfied, the resulting Shapley values can be interpreted in terms of relative and absolute importance. Using condition (iv) in addition to conditions (i)--(iii) also ensures that the Shapley values can be interpreted at a deeper structural level. For example, they can be interpreted as the fraction of the variance explained in linear regression when using $R^2$ as the goodness-of-fit measure.

Also, conditions (i)--(iii) are meaningful for regression models in general and are not limited to GLMs. Condition (iv) leads to a main focus of this paper, variable importance in GLMs.

\section{The Kullback-Leibler $R^2$ and its properties}
\label{R2KL}
In the previous section we emphasized that a careful choice of the goodness-of-fit measure is crucial, as it can avoid misinterpretation of results and also lead to an interpretation of Shapley values at a more structural level. 

It is assumed that we have a random sample $y_1, \dots, y_n$ from $f(y; \theta)$, a genuine (uncurved) one-parameter exponential family, with 
\begin{equation}
\label{eq:glm}
f(y; \theta) = \exp \left\{ y \theta - b(\theta) + c(y) \right\}, \; \theta \in \Theta,
\end{equation}
where $\theta$ is the natural or canonical parameter, $\Theta$ an interval of the real line, $b$ the cumulant function, and $c$ a function that does not depend on $\theta$. Different choices of the cumulant function $b$ lead to different models. In a GLM, the mean $\mu$ of $f$ is monotone in $\theta$ and is parameterized using a linear predictor $\eta_{i}$ and a known smooth and invertible link function $g$, with $g(\mu_{i}) = x_{i}^{\top}\beta = \eta_{i}$, 
where $x_{i} \in \mathbb{R}^{p}$ is the vector of the $p$ regressors for observation $i$ and $\beta \in \mathbb{R}^{p}$ is the vector of regression coefficients. Throughout it is assumed that models contain a constant term. Estimation is via maximum likelihood (ML); the maximum likelihood estimator (MLE) $\hat\theta$ is in the interior of $\Theta$. 

In a GLM setting, a unified approach to variable importance is possible when the Kullback-Leibler $R^{2}$, denoted as $R^{2}_{\text{KL}}$, is used as the goodness-of-fit measure. $R^{2}_{\text{KL}}$ was introduced by \citet{Cameron+Windmeijer:1997}. A main advantage of $R^{2}_{\text{KL}}$ is that it is meaningful for any regression model based on a (one-parameter) exponential family; see \citet{Cameron+Windmeijer:1997} and their Table~1 for an overview. Also, many well known (pseudo-) $R^{2}$ measures are special cases of $R^{2}_{\text{KL}}$, among them the classical $R^2$ for the linear regression model, $R^{2}_{\text{McF}}$ for binary response models, and the deviance $R^{2}$ for Poisson regression \citep{Cameron+Windmeijer:1996}. In addition, $R^{2}_{\text{KL}}$ 
can be interpreted in terms of the likelihood ratio test statistic (see Section \ref{R2KLi2}).

\subsection{The Kullback-Leibler $R^2$}
\label{sec:R2KL}

Recall that the Kullback-Leibler divergence \citep{Kullback+Leibler:1951} is defined as
\begin{equation}
\label{KLd}
K(\theta_{1}, \theta_{2} ) 
~:=~ 2 \; \mathsf{E}_{\theta_1} \left[ \log \left( \dfrac{ f(y, \theta_{1}) }{ f(y, \theta_{2}) } \right) \right] .
\end{equation}
It measures the information discrepancy between two densities, here represented by their parameters $\theta_{i}$, $i = 1,2$, using Shannon's entropy. $\mathsf{E}_{\theta_1}$ represents the expectation with respect to the model parameterized by $\theta_1$. For one-parameter exponential families,
\begin{equation}
\label{KLdexpfam}
K(\theta_{1}, \theta_{2} ) 
~=~ 2 \; \left[ (\theta_1 - \theta_2) \mu_1 - ( b(\theta_1) - b(\theta_2) ) \right] .
\end{equation}
Since the mean $\mu$ of $f$ is monotone in $\theta$, we can write $\mu = \mu(\theta)$ or $\theta = \theta(\mu)$, and also $K(\mu_{1}, \mu_{2})$ or $K(\theta_{1}, \theta_{2})$, as is convenient. In a GLM setting, if $\theta_1$ represents the saturated model with $\mu_1 = \mathbf{y}$, 
with $\mathbf{y}$ representing the data, we thus have 
\begin{equation*}
K(\mathbf{y}, \mu_{2} ) ~=~ 2 \; \left[ (\theta(\mathbf{y}) - \theta_2) \mathbf{y} - ( b(\theta(\mathbf{y})) - b(\theta_2) ) \right].
\end{equation*}
Genuine one-parameter exponential families are uncurved or flat \citep{Vos:1991} and the Kullback-Leibler divergence satisfies the Pythagorean relation \citep{Efron:1978, Hastie:1987},
\begin{equation*}
\label{pythag}
K(\hat\mu, \mu) ~=~ K(\mathbf{y}, \mu) - K(\mathbf{y}, \hat\mu) ,
\end{equation*}
where $\hat\mu$ represents the fitted model and $\mu$ some other model. 

For our purposes, $\mu = \hat \mu_0$, the model with only a constant term, is of particular interest. 
The corresponding Kullback-Leibler $R^{2}$ is now given by
\begin{equation}
\label{RDEV}
R^{2}_{\text{KL}} ~=~ 1 - \dfrac{K(\mathbf{ y}, \hat{\mu} )}{K(\mathbf{ y } , \hat{\mu}_{0} )} \quad \in [0,1].
\end{equation}
The idea is that the simplest model containing a constant only -- i.e., $\hat{\mu}_{0} = \bar{y}\mathbf{1}_{n}\in \mathbb{R}^{n}$ in the linear model, with $\mathbf{1}_{n}$ an $n$-vector of ones -- yields the maximum deviation from the `best' model, i.e., it maximizes the Kullback-Leibler divergence within a pre-specified model class. 
If a regressor contributes explanatory power to the model predictions $\hat{\mu}$ and leads to an improvement, we have $K(\mathbf{ y } , \hat{\mu} ) < K(\mathbf{ y } , \hat{\mu}_{0} )$, resulting in a larger $R^{2}_{\text{KL}}$ measure. The monotonicity condition from Sec.~\ref{ShvIntOv} is satisfied by construction, as the log-likelihood for a given model $\ell(\mathbf{y}, \hat{\mu})$ cannot decrease for an alternative model with an additional predictor. In view of equations (\ref{KLdiv2}), see below, and (\ref{RDEV}) the Kullback-Leibler $R^{2}$ satisfies the lower and upper bound conditions as it is bounded between zero and one; specifically, $R^{2}_{\text{KL}} = 0$ if $\hat{\mu} = \hat{\mu}_{0}$, and $R^{2}_{\text{KL}} = 1$ if $\hat{\mu} = \mathbf{y}$.



\cite{Cameron+Windmeijer:1997} outline that the Kullback-Leibler divergence can also be intuitively interpreted as an uncertainty measure. Specifically, the 
`deviation' of a fitted model from the optimal model corresponds to the empirical uncertainty, which can be measured by the Kullback-Leibler divergence employing the response and $\hat{\mu}$. 
%
%
Therefore, the importance measures in equations (\ref{ShvDecomp_imp}) and (\ref{ShvDecomp_impBest}) can be interpreted as follows:

\begin{itemize}
\item The Shapley value of predictor $i$ equals the fraction of the empirical uncertainty that is explained by predictor $i$, $\varphi_{i}$ = $impBM_{i}$, because $R^{2}_{\text{KL}}$ has an upper bound of one.

\item The contribution relative to the fitted model, $impFM_{i} =\varphi_{i} / v(P)$ equals the fraction of the explained empirical uncertainty that is explained by predictor $i$.
\end{itemize}

In summary, $v = R^{2}_{\text{KL}}$ is a convenient measure of fit, as it (a)~possesses the properties (i)--(iii) discussed in Section \ref{ShvIntOv}, (b)~is a generalization of the classical $R^{2}$ measure beyond linear regression, (c)~is based on established concepts from information theory, and (d)~is of a simple form for the widely used GLM class of models. 

The next subsection outlines a further interpretation in terms of the likelihood ratio test statistic. 

\subsection{The Kullback-Leibler $R^{2}$ and the likelihood ratio statistic}
\label{R2KLi2}

In regression modeling, it is desired that the coefficients of the fitted model with the vector of predictions $\hat{\mu}$ are jointly significant, compared to the model with only a constant term, i.e., with the vector of predictions $\hat{\mu}_0$. This can be tested using the likelihood ratio ($LR$) statistic,  
\begin{equation}
\label{LR}
LR ~=~ 2 \left( \ell(\mathbf{y}, \hat{\mu}) - \ell(\mathbf{y}, \hat{\mu}_{0}) \right).
\end{equation}
Using Hoeffding's representation of an exponential family \citep{Efron:1978, Hastie:1987}, the Kullback-Leibler divergence can also be expressed in terms of likelihoods, 
\begin{equation}
\label{KLdiv2}
K(\mathbf{ \mathbf{y}, \hat{\mu} }) ~=~ 2 \left( \ell(\mathbf{y}, \mathbf{y}) - \ell( \mathbf{y}, \hat{ \mu} ) \right),
\end{equation}
where $\ell(\mathbf{y}, \mathbf{y})$ is the log-likelihood of the saturated model and 
$\ell( \mathbf{y}, \hat{ \mu} )$ is the log-likelihood of a fitted model represented by $\hat{\mu} \in \mathbb{R}^{n}$. 

In GLM terminology, the quantity in equation (\ref{KLdiv2}) is the (residual) deviance; for GLMs, it plays the role of the residual sum of squares in the classical linear model. 
It follows that, for GLMs, the Kullback-Leibler $R^{2}$ is identical to the \emph{fraction of the deviance explained}.

From equations (\ref{KLdiv2}), (\ref{RDEV}) and (\ref{LR}) it is evident that $R^{2}_{\text{KL}}$ and $LR$ are closely related. In fact, in work preceding the Kullback-Leibler $R^{2}$, \citet{Magee:1990} already suggested to define fit measures via classical likelihood-based test statistics. 
Specifically, $R^{2}_{\text{KL}}$ is a scalar multiple of the likelihood ratio statistic \citep{Cameron+Windmeijer:1997}, 
\begin{align}
\label{RKLvsLR}
R^{2}_{\text{KL}} ~=~ \dfrac{1}{K(\mathbf{ y } , \hat{\mu}_{0} )} \; LR.
\end{align}
%

In view of the efficiency property we have $\sum_{i} \varphi_{i}(P, R^{2}_{\text{KL}}) = R^{2}_{\text{KL}}$, hence the Shapley values also correspond to a certain decomposition of the scaled likelihood ratio test statistic, 
\begin{align}
\label{RKLvsLR2}
K(\mathbf{ y } , \hat{\mu}_{0} ) \; \sum_{i} \varphi_{i}(P, R^{2}_{\text{KL}}) =  LR.
\end{align}
As all Shapley values are scaled by the same constant $K(\mathbf{ y } , \hat{\mu}_{0} )$, the null deviance, this leads to a further interpretation: The Shapley value for predictor $i$ can be interpreted as this predictor's contribution, up to a constant, to the overall likelihood ratio statistic of the model. 

%
%

\subsection{The Kullback-Leibler $R^{2}$ and McFadden's likelihood ratio index}
\label{R2McF_E}

As noted above, for binary response models a widely used pseudo-$R^2$ measure is McFadden's likelihood ratio index
\begin{equation}
R^{2}_{\text{McF}} ~=~ 1- \dfrac{\ell(\mathbf{y}, \hat{\mu})}{\ell(\mathbf{y}, \hat{\mu}_{0})} .
\end{equation}
Using equations (\ref{KLdiv2}) and (\ref{RDEV}), it follows that 
\begin{align}
\label{R2MFvsR2KL_app}
R^{2}_{\text{McF}} 
~=~ \left( 1 - \dfrac{\ell(\mathbf{y}, \mathbf{y})}{\ell(\mathbf{y}, \hat{\mu}_{0})} \right) R^{2}_{\text{KL}} = \zeta \, R^{2}_{\text{KL}}.
\end{align}
In the case of the Bernoulli distribution, where $y_i \in \{0,1\}$, 
\begin{align}
\ell(\mathbf{y}, \mathbf{y}) 
~=~ \sum_{i=1}^n \left( y_i \log(y_i) + (1-y_i) \log(1-y_i) \right) ~=~ 0 ,
\end{align}
implying $\zeta = 1$; hence $R^{2}_{\text{McF}} = R^{2}_{\text{KL}}$ in the binary response case. However, beyond that case one generally has $\zeta \neq 1$, which results in $R^{2}_{\text{McF}}$ violating the upper bound condition. We illustrate this issue in the following section using a Poisson model.

\section{Examples}
\label{ShvExample}

This section provides examples of assessing variable importance using Shapley values for selected GLMs (and certain extensions thereof). 
Subsequently, we denote by $\ell(S)$ and $R^2_{\text{KL}}(S)$ the goodness of fit measures computed for the regression model containing the predictors $S\subseteq P$, where the cardinality of $P$ is equal to $p$. 
For example, $\ell(\varnothing)$ corresponds to the log-likelihood of the model with only a constant term. 
Similarly, $\hat{\mu}_S$ is the vector of predictions calculated from the regression model using the set of predictors $S$. 

The following points are emphasized:
\begin{enumerate}
\item Interpretation of Shapley values in terms of relative and absolute importance.
\item Consequences of violations of the lower and upper bound conditions.
\item Relations among Shapley values derived from different likelihood-based quantities.
\end{enumerate}

\subsection{Poisson regression}
\label{illustration_poisson}

We begin with Poisson regression, based on the probability density
\begin{equation}
\label{dpoiss}
f(y_{i} ; x_{i}, \mu_{i}) ~=~ \dfrac{e^{-\mu_{i}} \mu_{i}^{y_{i}}}{y_{i}!}, \quad \text{for}~y_{i} \in \mathbb{N}_{0}, 
\end{equation}
where $\mu_{i} = \mathsf{E}(y_{i} \mid x_{i})$ and $\eta_{i} = \log(\mu_i)$. 


We use data from health economics that were originally analyzed by \citet{Cameron+Trivedi:1986}, see also \citet{Cameron+Trivedi:2013}, and which are available from the data archive of the \textit{Journal of Applied Econometrics}\footnote{\url{http://qed.econ.queensu.ca/jae/1997-v12.3/mullahy/}}. For \textsf{R} users, they are also available under the name \texttt{DoctorVisits} from the \textsf{R}~package \textbf{AER} \citep{Kleiber+Zeileis:2008}, where a detailed description of all variables can be found. The response is the number of doctor visits, \texttt{visits}, with a maximum count of $9$. 
The regressors provide information on the health status and on socioeconomic characteristics of the patients in the sample. More specifically, we make use of the variables \texttt{age}, \texttt{gender}, \texttt{health}, \texttt{illness}, \texttt{income}, \texttt{lchronic}, \texttt{nchronic}, \texttt{private} and \texttt{reduced}. The rootogram \citep{Kleiber+Zeileis:2016} of the model using this set of regressors confirms that the Poisson model is a suitable choice for these data; it is shown in Figure~\ref{root_poi}. 

\begin{figure}[t!]
\centering
\includegraphics{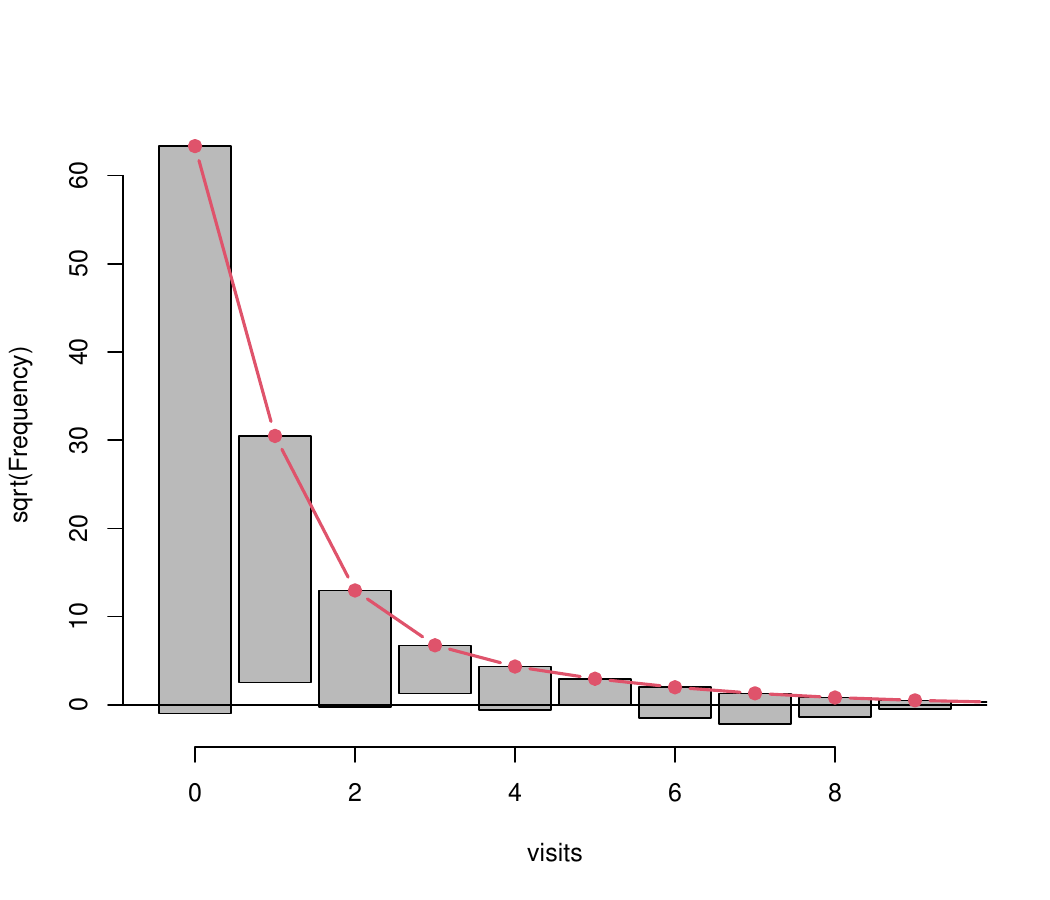}
\caption{\label{root_poi} Rootogram of the Poisson regression model using all regressors.}
\end{figure}

The Shapley values are obtained using the Kullback-Leibler $R^{2}$, here given by   
\begin{equation}
\label{R2DP}
R^{2}_{\text{KL,Poi}} 
~=~ 1 - \dfrac{\sum_{i=1}^{n} \big[y_{i} \log(y_i / \hat{\mu}_{i}) - (y_i - \hat{\mu}_{i})\big]}{\sum_{i=1}^{n} y_{i} \log(y_{i}/\bar{y})} ,
\end{equation}
which equals the Poisson deviance $R^{2}$ proposed by \citet{Cameron+Windmeijer:1996}. To illustrate the consequences of violating the requirements discussed in Section \ref{ShvIntOv} we also use McFadden's $R^{2}$. 
The latter is implemented, for example, in the \textsf{R}~package \textbf{dominanceanalysis} for use with several GLMs, including Poisson regression.

From equation (\ref{R2MFvsR2KL_app}) we know that for all regression models with $\ell(\mathbf{y}, \mathbf{y}) \neq 0$, i.e., $\zeta \neq 1$, it holds that $R^{2}_{\text{McF}} \neq R^{2}_{\text{KL}}$. For our data and the chosen Poisson model we have $\zeta = 0.71$, hence the range of values of $R^{2}_{\text{McF}}$ is smaller than that of $R^{2}_{\text{KL}}$, which is the unit interval. This results in the violation of the upper bound condition for $R^{2}_{\text{McF}}$ for this Poisson model. Appendix \ref{relR2KLell} shows that for the Shapley values we also have $\varphi_i(P,R^2_{\text{McF}}) = \zeta \, \varphi_i(P,R^2_{\text{KL}})$.\footnote{Any numerical deviations from this relationship occurring in Table \ref{tab:Pois_Reg} are due to rounding.} 
Therefore, the Shapley values based on $R^{2}_{\text{McF}}$ can no longer be interpreted relative to the best model, unless they are rescaled by $\zeta$. 

\begin{table}[ht]
\centering
\caption{Largest five Shapley values for the Poisson regression model, 
            using $v = R^{2}_{\text{KL}}$ and $v = R^{2}_{\text{McF}}$,
            respectively.} 
\label{tab:Pois_Reg}
\begin{tabular}{lrrrrrr}
  \\[-1.8ex] \hline 
 \hline \\[-1.8ex] 
 & reduced & illness & health & lchronic & age & $v(P)$ \\ 
  \hline \\[-1.8ex] 
$v = R^{2}_{\text{KL}}$ & 0.1250 & 0.0415 & 0.0207 & 0.0111 & 0.0110 & 0.2211 \\ 
  $v = R^{2}_{\text{McF}}$ & 0.0884 & 0.0293 & 0.0146 & 0.0079 & 0.0078 & 0.1564 \\ 
   \hline 
\end{tabular}
\end{table}
We illustrate the issues in Table \ref{tab:Pois_Reg}. The empirical uncertainty explained by the model is about $22.11\%$. The largest contribution comes from the predictor \texttt{reduced};  
its relative importance, $impFM_{\texttt{reduced}}$, is about $56.54\%$. 
Moreover, as the saturated model has a goodness-of-fit of unity for $R^{2}_{\text{KL}}$, the Shapley values can also be interpreted as importance measures relative to the best model (see Section \ref{UpperBound}). 
Thus, from $\varphi_{\texttt{reduced}}(P, R^{2}_{\text{KL}}) = impBM_{\texttt{reduced}} = 12.5\%$ we see that the predictor \texttt{reduced} explains $12.5\%$ of the total empirical uncertainty and is therefore of considerable relevance for explaining the number of doctor visits. In contrast, the importance assessments using $v = R^{2}_{\text{McF}}$ 
do not allow such interpretations without explicitly calculating $\zeta$, because, as noted above, $R^{2}_{\text{McF}}$ violates the upper bound condition in this application. This illustrates that special care should be taken when choosing the goodness-of-fit measure, as its choice affects the interpretability of importance measures.

We add that since $R^{2}_{\text{KL}}$ corresponds to the scaled likelihood ratio test statistic, 
the Shapley values from $R^{2}_{\text{KL}}$ can also be interpreted as contributions to this test statistic. In our case, \texttt{reduced} is the predictor with the largest contribution. 

\subsection{Poisson hurdle regression}	
\label{HPM}

In the area of count data regression, many data sets are plagued by a large number of zero observations, so that classical models such as the Poisson model do not provide an adequate fit. A more flexible specification is the hurdle regression model, also known as a two-part model, originally proposed by \citet{Mullahy:1986}. More formally, the hurdle model is a combination of two models, $f_{1}$ and $f_{2}$, where $f_{1}$ represents a binary response part, often of logit form, and $f_{2}$ is a count data model that is left-truncated at $y = 1$. Overall, the hurdle model is given by
\begin{align}
\label{dhurdle}
f(y_{i}; x_{i1},x_{i2}, \tau_{1}, \tau_{2}) 
~=~  
\begin{cases} 
f_{1}(0;x_{i1},\tau_{1}), 
 & \text{for}~y_{i}=0 , \\ 
\dfrac{1-f_{1}(0;x_{i1},\tau_{1})}{1-f_{2}(0;x_{i2},\tau_{2})} f_{2}(y_{i};x_{i2},\tau_{2}), 
 & \text{for}~y_{i}>0 , 
\end{cases}
\end{align}
where $x_{i1}, x_{i2}$ are the vectors of regressors for observation $i$ and $\tau_{1}, \tau_{2}$ are the corresponding vectors of regression coefficients, respectively. In general, $x_{i1} \neq x_{i2}$, but specifications where $x_{i1} = x_{i2}$ are quite common in the empirical literature. 

The log-likelihood of the hurdle model, $\ell_{\text{hurdle}}$, can be split into two components \citep{Mullahy:1986}: the log-likelihood of a binary response model, $\ell_{\text{binary}}$, 
and the log-likelihood of a zero-truncated count regression model, $\ell_{\text{zt-count}}$; i.e., $\ell_{\text{hurdle}} ~=~ \ell_{\text{binary}} + \ell_{\text{zt-count}}$. This implies that the hurdle model can be estimated by fitting both parts separately, and that variable importance can be assessed part by part.

For the binary part, we use a logit model, with $y^*_i$ an indicator of positive counts, 
\begin{equation}
\label{logitmodel}
f(y^*_{i}; x_{i1}, \mu_{i1}) 
~=~ \mu_{i1}^{y^*_{i}} \, (1 - \mu_{i1})^{(1 - y^*_{i})}, \quad \text{for} ~ y^*_{i} \in \{0, 1\},
\end{equation}
where $\mu_{i1} = \mathsf{P}(y^*_{i}=1 \mid x_{i 1}) = \mathsf{E}(y^*_{i} \mid x_{i1})$ and $\eta_{i} = \log\left(\mu_{i1}/(1-\mu_{i1}) \right)$.

For the positives, we use a zero-truncated Poisson model, with 
\begin{equation}
\label{ztpmodel}
f(y_{i} ; x_{i2}, \mu_{i2}) 
~=~ \dfrac{\mu_{i2}^{y_{i}}}{(e^{\mu_{i2}} - 1) \; y_{i}! }, \quad \text{for}~y_{i} \in \mathbb{N},
\end{equation}
where $\mu_{i2} = \mathsf{E}(y_{i} \mid x_{i2})$ and $\eta_{i} = \log(\mu_{i2})$. 
The zero-truncated Poisson distribution is still an exponential family, hence it naturally leads to a GLM.

The two log-likelihood components for our model are thus given by
\begin{align}
\nonumber 
\ell_{\text{logit}}(\mathbf{y}^*, \mu_1)  
&= \sum_{i=1}^n \left[ y^*_i \log(\mu_{i1}) + (1-y^*_i) \log(1-\mu_{i1}) \right] , \\
\ell_{\text{ztPoi}}(\mathbf{y}, \mu_2) 
&= \sum_{\{i:y_{i}>0\}} \left[y_i \log(\mu_{i2}) - \log\left(e^{\mu_{i2}} - 1 \right) - \log(y_i!) \right].
\label{eq:hurdle_loglik}
\end{align}
Overall, equations (\ref{eq:hurdle_loglik}) imply that the Poisson hurdle model has two GLM building blocks, a logit model for the binary part and a zero-truncated Poisson regression model for positive counts. They also imply that variable importance can be assessed by using the Shapley value approach separately for each of these two building blocks.

\begin{figure}[t!]
\centering
\includegraphics{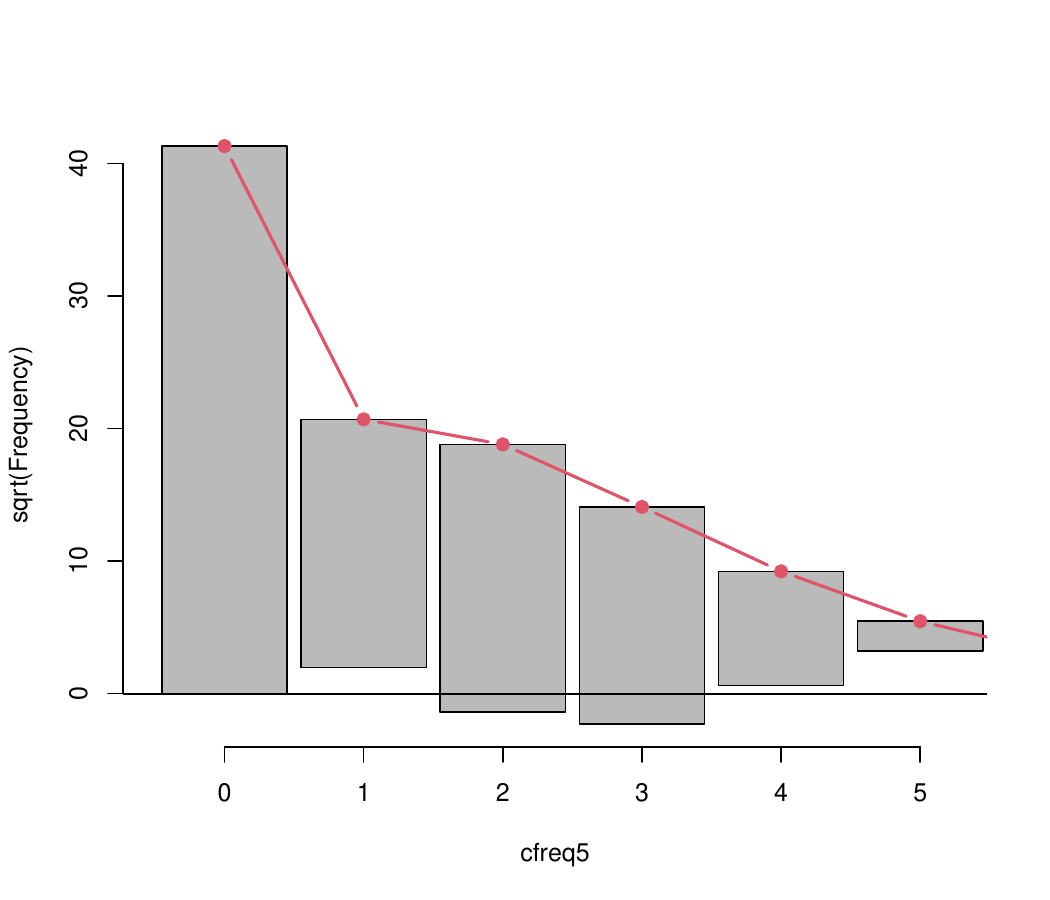}
\caption{\label{root_hpoi} Rootogram of the Poisson hurdle model using all 13 regressors.}
\end{figure}

We use this Poisson hurdle model to model car insurance data previously analyzed by \citet{Yip+Yau:2005}. Their data are available via the \texttt{AutoClaim} data from the \textsf{R}~package \textbf{cplm} \citep{Zhang:2013}. Specifically, the relevant subset of observations can be extracted via a binary factor, \verb+IN_YY+, indicating inclusion in the Yip and Yau paper. 
The response variable of interest is the number of claims in the past five years, \texttt{cfreq5}, with a maximum count of $5$. Yip and Yau use five regressors in their main analysis, but start out from a larger data set of 13 regressors, which in turn are taken from an even larger data set. 
Given our interest in variable importance we use their initial set of 13 regressors: \texttt{age}, \texttt{area}, \texttt{cartype}, \texttt{educ}, \texttt{gender}, \texttt{income}, \texttt{jobclass}, \texttt{married}, \texttt{red}, \texttt{revoked}, \texttt{singlep}, \texttt{usage} and \texttt{violation}.\footnote{The original names of the variables in the \texttt{AutoClaim} data are, in the same order: AGE, AREA, CAR\_TYPE, MAX\_EDUC, GENDER, INCOME/10000, JOBCLASS, MARRIED, RED\_CAR, REVOLKED, PARENT1, CAR\_USE and MVR\_PTS. Note that \texttt{income} is INCOME scaled by 10000. The response \texttt{cfreq5} was originally called CLM\_FREQ5.} See \citet{Yip+Yau:2005} and the documentation of the \texttt{AutoClaim} data for further information on these predictors. The rootogram in Figure~\ref{root_hpoi} confirms that the claim frequency variable is adequately modelled by a Poisson hurdle model using these predictors.

The Kullback-Leibler $R^{2}$ for the zero-truncated Poisson model, $R^{2}_{\text{KL,ztPoi}}$, is given by
\begin{align}
\label{R2DZP}
R^{2}_{\text{KL, ztPoi}} 
&= 1 - \dfrac{\sum_{\{i:y_{i}>0\}} \big[ y_{i} \log(y_{i}/\hat{\mu}_{i}) - \log(\exp(y_{i})-1) + \log(\exp(\hat{\mu}_{i}) - 1)\big]}{\sum_{\{i:y_{i}>0\}} \big[y_{i}\log(y_{i}/\bar{y}_{+})-\log(\exp(y_{i})-1) + \log(\exp(\bar{y}_{+}) - 1)\big]},
\end{align}
where $\bar{y}_{+} = n_{+}^{-1} \sum_{\{i:y_{i}>0\}} y_{i}$, with $n_{+}$ corresponding to the number of positive responses ($y_i>0$). For the binary part with a logit link we have \citep{Cameron+Windmeijer:1997}
\begin{align}
\label{R2MF}
R^{2}_{\text{KL, logit}} 
= 1 - \dfrac{ \sum_{i=1}^{n} \big[ y^*_{i} \log( \hat{\mu}_{i} ) + (1-y^*_{i}) \log( 1- \hat{\mu}_{i}) \big]}{  \sum_{i=1}^{n} \big[ y^*_{i} \log( \bar{y}^* ) + (1-y^*_{i}) \log( 1- \bar{y}^* )\big] }.
\end{align}
As noted above, the resulting Kullback-Leibler $R^{2}$ for the binary part is identical to $R^{2}_{\text{McF}}$.

\begin{table}[ht]
\centering
\caption{Largest five Shapley- and pseudo-Shapley values for the positive part of the hurdle model, using $v = R^{2}_{\text{KL}}$, 
            $v = \ell - \ell(\varnothing)$ and $v^{*} = \ell$, respectively.} 
\label{tab:Hurdle_ztPois_Reg}
\begin{tabular}{lrrrrrr}
  \\[-1.8ex] \hline 
 \hline \\[-1.8ex] 
 & cartype & jobclass & red & educ & singlep & $v(P)$ or $v^{*}(P)$ \\ 
  \hline \\[-1.8ex] 
$v = R^{2}_{\text{KL}}$ & 0.0103 & 0.0067 & 0.0045 & 0.0036 & 0.0036 & 0.0340 \\ 
  $v = \ell - \ell(\varnothing)$ & 2.9199 & 1.9136 & 1.2756 & 1.0252 & 1.0212 & 9.6376 \\ 
  $v^{*} = \ell$ & 2.9199 & 1.9136 & 1.2756 & 1.0252 & 1.0212 & -1453.3342 \\ 
   \hline 
\end{tabular}
\end{table}
Table~\ref{tab:Hurdle_ztPois_Reg} provides the five largest Shapley values for the positive part of the hurdle model using all 13 variables, along with the resulting goodness-of-fit measures $R^{2}_{\text{KL}}$, the log-likelihood, and the shifted log-likelihood. (The choice of fit measures is motivated by their availability in software implementations, specifically in the \textsf{R}~packages \textbf{hier.part} and \textbf{dominanceanalysis}.) The table nicely illustrates the problems arising from the violation of the lower and upper bound conditions, as discussed in Sections \ref{LowerBound} and \ref{UpperBound}. 
First, the pseudo-Shapley values associated with the log-likelihood (bottom row in Table~\ref{tab:Hurdle_ztPois_Reg}) might suggest some relevance of the variables, although the overall fit in terms of $R^{2}_{\text{KL}}$ is quite poor, as only about $3.4\%$ of the empirical uncertainty is explained by the model. The value of the corresponding overall log-likelihood $v^{*}(P) = -1453.33$ does not provide much useful information here, it is negative and difficult to interpret. 
Second, the efficiency condition is satisfied for the shifted log-likelihood $v = \ell - \ell(\varnothing)$, and the corresponding overall value $v(P) = 9.64$ might suggest a good model (recall that a value of zero corresponds to the worst model). However, there is still a lack of interpretability, as the fit of the best model is not known or computed. The missing reference point of the best model does not allow to assess whether $v(P) = 9.64$ is `small' or `large'.
This problem does not arise with a fit measure that satisfies the upper and lower bound conditions, such as $R^{2}_{\text{KL}}$, where the value $v(P)=0.034$ can be compared against the built-in upper bound of 1. Using $R^{2}_{\text{KL}}$, the Shapley values allow, in addition to an interpretation in terms of relative importance, an interpretation in terms of absolute importance. For example, \texttt{cartype} is about $1.54$ times more important than \texttt{jobclass}, while in absolute terms both predictors are not important.

\begin{table}[ht]
\centering
\caption{Largest five Shapley- and pseudo-Shapley values for the binary part of the hurdle model, using $v = R^{2}_{\text{KL}}$,
            $v = \ell - \ell(\varnothing)$ and $v^{*} = \ell$, respectively.} 
\label{tab:Hurdle_logit_Reg}
\begin{tabular}{lrrrrrr}
  \\[-1.8ex] \hline 
 \hline \\[-1.8ex] 
 & violation & area & cartype & jobclass & educ & $v(P)$ or $v^{*}(P)$ \\ 
  \hline \\[-1.8ex] 
$v = R^{2}_{\text{KL}}$ & 0.1612 & 0.0580 & 0.0038 & 0.0034 & 0.0024 & 0.2353 \\ 
  $v = \ell - \ell(\varnothing)$ & 303.7332 & 109.3837 & 7.1795 & 6.4305 & 4.4890 & 443.5264 \\ 
  $v^{*} = \ell$ & 303.7332 & 109.3837 & 7.1795 & 6.4305 & 4.4890 & -1441.0973 \\ 
   \hline 
\end{tabular}
\end{table}
The results for the binary part are summarized in Table~\ref{tab:Hurdle_logit_Reg}. As before, the violations of the lower and upper bound conditions lead to problems of interpretation. For $v^* = \ell$, an ordering of the predictors in terms of relative importance is possible, but statements about the importance relative to the fitted ($impFM$) and `best' ($impBM$) models are impossible without knowing $\ell(\varnothing)$. For $v = \ell - \ell(\varnothing)$, statements about $impFM$ become feasible. The Shapley values for \texttt{violation} and \texttt{area} are the largest -- these regressors explain 68.5\% and 24.7\% of the fitted model, respectively --, while the other predictors appear to be much less relevant. Using $v = R^{2}_{\text{KL}}$ provides further insight, as the Shapley values can now be interpreted as absolute importance measures. The predictors \texttt{violation} and \texttt{area} explain 16.12\% and 5.80\% , respectively, relative to the best model. 

A comparison of Tables~\ref{tab:Hurdle_ztPois_Reg} and~\ref{tab:Hurdle_logit_Reg} reveals another interesting detail: the binary response model explains $23.53\%$ of the empirical uncertainty, whereas the zero-truncated Poisson model in Table~\ref{tab:Hurdle_ztPois_Reg} explains only $3.40\%$. Therefore, the binary part of this two-part model performs much better than the zero-truncated Poisson part. In practical terms, this means that while the available regressors are useful for modelling claim incidence, they are of limited relevance for modelling the exact number of claims.

Furthermore, since $R^{2}_{\text{KL}}$ is a likelihood-based goodness-of-fit measure, the `distortions' of the (pseudo-) Shapley values resulting from the use of $v = \ell - \ell(\varnothing)$ and $v^* = \ell$ can be quantified. All three goodness-of-fit measures are linearly related, in particular it can be shown (see Appendix~\ref{relR2KLell}) that 
\begin{align}
\varphi_i(P,R^{2}_{\text{KL}}) 
~=~  \underbrace{\dfrac{1}{ \ell(\mathbf{y},\mathbf{y}) - \ell(\mathbf{y},\hat{\mu}_{0}) }}_{=:C} \, \varphi^{*}_{i}(P, \ell) 
~=~  \underbrace{\dfrac{1}{ \ell(\mathbf{y},\mathbf{y}) - \ell(\mathbf{y},\hat{\mu}_{0}) }}_{=:C} \, \varphi_{i}(P, \ell-\ell(\varnothing)) , 
\label{eq:normalisation}
\end{align}
where $C$ is the null deviance. 
In other words, the resulting (pseudo-) Shapley values inherit the linear relationship that exists between the fit measures.  For the data at hand, 
we have $C^{-1} = 283.75$ for the zero-truncated Poisson model and $C^{-1} = 1884.62$ for the logit model, respectively.\footnote{The values of $C$ calculated from Tables~\ref{tab:Hurdle_ztPois_Reg} and \ref{tab:Hurdle_logit_Reg} differ slightly due to rounding.} 

\subsection{Geometric regression}
\label{illustration_geometric}

\begin{figure}[t!]
\centering
\includegraphics{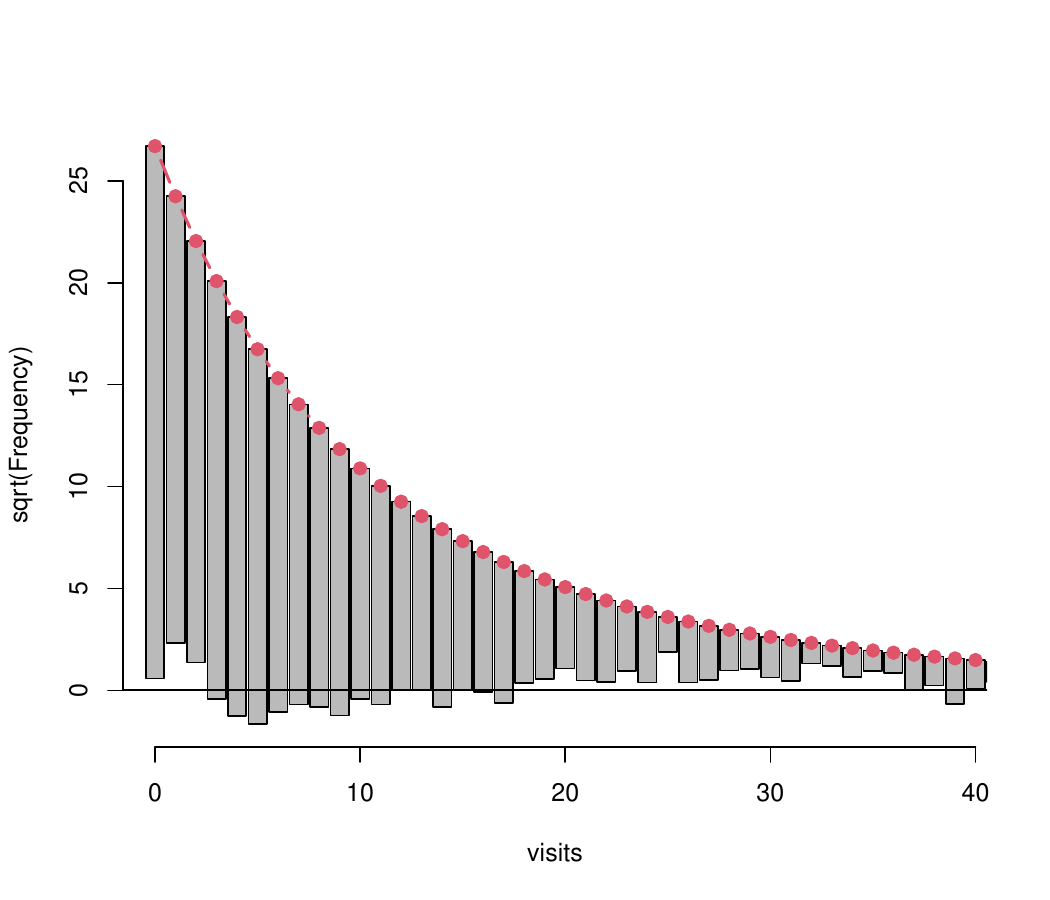}
\caption{\label{root_geom} Rootogram of the geometric regression model using all regressors.}
\end{figure}

Another typical problem with count data is overdispersion, i.e., the presence of more variability in a data set than would be expected based on a given model for the mean (the implicit reference point being the Poisson model).  
Our final example, therefore, presents a count data model that allows for a substantial amount of overdispersion. Again, we use data from health economics, originally analyzed by \citet{Deb+Trivedi:1997}, see also \citet{Cameron+Trivedi:2013}. These data are also available from the data archive of the \textit{Journal of Applied Econometrics}\footnote{\url{http://qed.econ.queensu.ca/jae/1997-v12.3/deb-trivedi/}}. For \textsf{R}~users, they are furthermore available under the name \texttt{NMES1988} from the \textsf{R}~package \textbf{AER} \citep{Kleiber+Zeileis:2008}, where a detailed description of all variables can be found. The response is the number of physician office visits, \texttt{visits}, with a maximum count of $89$. The regressors \texttt{adl}, \texttt{afam}, \texttt{age}, \texttt{chronic}, \texttt{employed}, \texttt{gender}, \texttt{health}, \texttt{income}, \texttt{insurance} and \texttt{married} provide information on the health and the socioeconomic status of the sample persons. 

Figure~\ref{root_geom} shows the rootogram for the model using all regressors, confirming that the geometric regression model is a suitable choice for these data. We therefore use 
\begin{align}
\label{dgeometric}
f(y_{i} ; x_{i}, \mu_{i}) 
~=~ \dfrac{ \mu_{i}^{y_{i}} }{(1+\mu_{i})^{(y_{i}+1)}}, ~~\text{for}~y_{i} \in \mathbb{N}_{0},
\end{align}
where $\mu_{i} = \mathsf{E}(y_{i} \mid x_{i})$ and $\eta_{i} ~=~ \log (\mu_{i})$. 
As is common with count data, a log link is used, although this is not the canonical link for the model at hand. 

The Shapley values are obtained using $R^{2}_{\text{KL}}$, here of the form \citep{Cameron+Windmeijer:1997}
\begin{align}
R^{2}_{\text{KL,geo}} 
&= 1 - \dfrac{ \sum_{i=1}^{n} \left[y_{i} \log\left( y_{i}/\hat{\mu}_{i} \right) - (1+y_{i}) \log\left( (1 + y_{i})/(1+\hat{\mu}_{i}) \right)\right]}{\sum_{i=1}^{n} \left[y_{i} \log\left(y_{i}/\bar{y} \right) - (1+y_{i}) \log\left( (1 + y_{i})/(1+\bar{y}) \right)\right]}.
\end{align}

\begin{table}[ht]
\centering
\caption{Largest five Shapley values for the geometric regression model, 
           for $v = R^{2}_{\text{KL}}$.} 
\label{tab:Geom_Reg}
\begin{tabular}{lrrrrrr}
  \\[-1.8ex] \hline 
 \hline \\[-1.8ex] 
 & chronic & health & insurance & adl & afam & $v(P)$ \\ 
  \hline \\[-1.8ex] 
$v = R^{2}_{\text{KL}}$ & 0.0535 & 0.0233 & 0.0086 & 0.0056 & 0.0017 & 0.0953 \\ 
   \hline 
\end{tabular}
\end{table}
Table \ref{tab:Geom_Reg} gives the Shapley values based on $R^{2}_{\text{KL}}$. We refrain from comparing this with alternative measures of fit, as we are not aware of empirical work using variable importance measures in conjunction with geometric regression. The predictor \texttt{chronic}, which corresponds to the number of chronic conditions, has the largest contribution, followed by \texttt{health}. However, while \texttt{chronic} has a large relative importance within the fitted model, namely $impFM_{\texttt{chronic}} = 56.14\%$, the explanatory power of the full model is not impressive, with $R^{2}_{\text{KL}}$ approximately equal to $9.53\%$. The absolute importance of \texttt{chronic} 
is about $5.35\%$. Thus, the regressor \texttt{chronic} explains only $5.35\%$ of the total empirical uncertainty and therefore seems to have limited explanatory power regarding the number of physician office visits. This example again highlights the usefulness of goodness-of-fit measures that have the properties from Section \ref{ShvIntOv} and thus allow interpretation in terms of absolute importance.

\section{Conclusion}
\label{conclusion}

Understanding the importance of explanatory variables in regression models is of central interest in many fields. One popular approach is based on the Shapley value, a concept originating from game theory. A key component in calculating the Shapley value is the characteristic function or, in regression terminology, a suitable goodness-of-fit measure. In statistical literature, this idea has primarily been applied to linear regression models, for which the classical $R^{2}$ is a natural starting point. In this context, the Shapley values offer a `fair' decomposition of the classical $R^{2}$. 

However, there is currently no widely accepted framework for evaluating variable importance in GLMs. We present a unified approach for GLMs, building on previous contributions for linear regression and for binary response models. We also present and discuss desirable properties of goodness-of-fit measures, some of which apply to regression models beyond GLMs. We demonstrate that these properties enable Shapley values to be interpreted as measures of relative and absolute importance. 
Furthermore, we propose using the Kullback-Leibler $R^{2}$, which, for GLMs, is identical to the fraction of deviance explained 
and generalizes several well-known fit measures, such as the classical $R^2$ and McFadden's likelihood ratio index for binary response models. 

The Kullback-Leibler $R^2$ may also be the goodness-of-fit measure of choice for several nonlinear regression models that are not based on distributions that form an exponential family. This is currently under investigation. However, the present paper takes the first steps beyond the GLM framework by using a Poisson hurdle model in Section~\ref{ShvExample}. This model has GLM building blocks, but its two-part structure makes it more flexible than a GLM.


\bibliographystyle{agsm}
\bibliography{p_shv_varimp}

\newpage

\begin{appendix}

\section{Relationships among goodness-of-fit measures and \newline (pseudo-) Shapley values}
\label{relR2KLell}

This appendix outlines the relationships among four goodness-of-fit measures, $R_{\text{KL}}^{2}$, $R_{\text{McF}}^{2}$, $\ell - \ell(\varnothing)$ and $\ell$, and the corresponding (pseudo-) Shapley values. 

First, as these quantities are all linearly related, we can study the relationships in one go. 
Specifically, consider a set of regressors $P$ and two linearly related goodness-of-fit measures, 
$v_2(S) = a \, v_1(S) + b$, where $a, \, b \in \mathbb{R}$ and $S \subseteq P$ is some subset of the available regressors. 
In view of equation (\ref{shapleyvalue}) the shift term $b$ drops out, so the resulting \mbox{(pseudo-) }Shapley values are proportional to each other: namely, $\varphi_i(P, v_2) = a \, \varphi_i(P, v_1)$. Note also that if a fit measure $v$ does not satisfy $v(\varnothing) = 0$, we obtain pseudo-Shapley values $\varphi^*_i(P, v)$.

From equations (\ref{KLdiv2}) and (\ref{RDEV}) we can write, for given data $\mathbf{y}$ and a subset $S$ of regressors leading to predictions $\hat{\mu}_S$, 
\begin{equation}
R^2_{\text{KL}}(\hat{\mu}_S) 
~=~ \frac{ \ell(\mathbf{y}, \hat{\mu}_S) - \ell(\mathbf{y},\hat{\mu}_{0}) }{ \ell(\mathbf{y}, \mathbf{y}) - \ell(\mathbf{y}, \hat{\mu}_{0}) },
\label{eq:KL_devianceform}
\end{equation}
where $\ell(\mathbf{y}, \mathbf{y})$ represents the log-likelihood of the saturated model. 
As before, it is convenient to express this identity in terms of the relevant subset $S$, by using $R^2_{\text{KL}}(S) = R^2_{\text{KL}}(\hat{\mu}_S)$ and $\ell(S) = \ell(\mathbf{y}, \hat{\mu}_S)$, with $S\subseteq P$. 
As in Section \ref{UpperBound}, let $P'$ denote the saturated model. Then, equation~(\ref{eq:KL_devianceform}) can be reformulated as 
\begin{equation}
R^2_{\text{KL}}(S) 
~=~ \frac{1}{ \ell(P') - \ell(\varnothing) } \, \ell(S) 
 - \frac{\ell(\varnothing)}{\ell(P') - \ell(\varnothing)} ;
\label{eq:relship_KL_la}
\end{equation}
it follows that
\begin{equation}
\varphi_i(P,R^{2}_{\text{KL}}) 
~=~ \frac{1}{\ell(P') - \ell(\varnothing) } \, \varphi^{*}_{i}(P, \ell) . 
\label{eq:relship_KL_lb}
\end{equation}
Similarly, equation (\ref{R2MFvsR2KL_app}), reformulated in terms of sets of regressors, leads to
\begin{equation}
R^{2}_{\text{KL}}(S)
~=~ \frac{-\ell(\varnothing)}{\ell(P')-\ell(\varnothing)} \, R^{2}_{\text{McF}}(S) ; \label{eq:relship_KL_McFa}
\end{equation}
it follows that
\begin{equation}
\varphi_i(P,R^{2}_{\text{KL}}) 
~=~ \frac{-\ell(\varnothing)}{ \ell(P') - \ell(\varnothing) } \, \varphi_{i}(P, R^{2}_{\text{McF}}). \label{eq:relship_KL_McFb}
\end{equation}
Finally, the relation between $\ell(S) - \ell(\varnothing)$ and $\ell(S)$ is obviously linear, with $a=1$, hence
\begin{equation}
\varphi_i(P,\ell - \ell(\varnothing)) ~=~ \varphi^*_{i}(P, \ell).
\end{equation}
Note that using $R_{\text{KL}}^{2}(S)$, $R_{\text{McF}}^{2}(S)$ or $\ell - \ell(\varnothing)$ results in Shapley values, whereas using $\ell(S)$ results in pseudo-Shapley values. 

A goodness-of-fit measure is monotonically increasing with respect to the addition of a new regressor. This results in a positive multiplicative constant in equation (\ref{eq:relship_KL_la}), hence a mapping from and to positive Shapley values. The sign of the multiplicative constant in equation (\ref{eq:relship_KL_McFa}) is less obvious, because the sign of the log-likelihood depends on the model and the data at hand. 
However, for binary response models the log-likelihood is always nonpositive; therefore, the multiplicative constant in equation (\ref{eq:relship_KL_McFa}) is nonnegative for these models. (In fact, it is equal to one, since McFadden's likelihood ratio index, $R^{2}_{\text{McF}}$, is equal to $R^{2}_{\text{KL}}$ in this case.)

\end{appendix}

\end{document}